\documentclass[12pt, a4paper]{article}
\usepackage[font=footnotesize]{caption}
\usepackage[utf8]{inputenc}
\usepackage{graphicx}
\usepackage{authblk}
\usepackage{booktabs}
\usepackage{verbatim}
\usepackage{url}
\usepackage{caption}
\usepackage{subcaption}
\usepackage{indentfirst}
\usepackage{url}
\usepackage{amssymb}
\usepackage{amsmath}
\title{Parallel and Sequential Resources Networks}
\author{Alexandre Benatti$^1$ and Luciano da F. Costa$^2$}

\affil{
$^1$Institute of Mathematics and Statistics - DCC \\
University of S\~ao Paulo \\
Rua do Mat\~ao, 1010, S\~ao Paulo, SP 05508-090 Brazil 
\\ \vspace{0.5cm}
$^2$S\~ao Carlos Institute of Physics - DFCM \\
University of S\~ao Paulo \\
Av. Trabalhador S\~ao-Carlense, 400, S\~ao Carlos, SP 13566-590 Brazil
}

\date{\today}

\begin{document}

\maketitle

\begin{abstract}
A large number of real and abstract systems involve the transformation of some basic resource into respective products under the action of multiple processing agents, which can be understood as multiple-agent production systems (MAP). At each discrete time instant, for each agent, a fraction of the resources is assumed to be kept, forwarded to other agents, or converted into work with some efficiency. The present work describes a systematic study of nine basic MAP architectures subdivided into two main groups, namely parallel and sequential distribution of resources from a single respective source. Several types of interconnections among the involved processing agents are also considered. The resulting MAP architectures are studied in terms of the total amount of work, the dispersion of the resources (states) among the agents, and the transition times from the start of operation until the respective steady state. Several interesting results are obtained and discussed, including the observation that some of the parallel designs were able to yield maximum work and minimum state dispersion, achieved at the expense of the transition time and use of several interconnections between the source and the agents. The results obtained for the sequential designs indicate that relatively high performance can be obtained for some specific cases.
\end{abstract}

\section{Introduction}

A wide range of natural and artificial systems can be modeled in terms of multiple agents transforming some resource into useful results and/or products, which will be henceforth referred to as \emph{multi-agent production} (e.g.~\cite{kutanoglu2007coalitions,behnamian2012incorporating,giordani2013distributed,may2021decentralized}), or \emph{MAP} for short. For simplicity's sake, the present work focuses on resources emanating from a single source. 

Examples of MAP systems include, but are by no means restricted to: (i) multi-cellular living beings maintaining their life relying on nourishment sources; (ii) individual production lines in which raw materials (resources) are transformed into products; (c) distribution of resources among production plants; and (d) information processing by a set of computing elements. The ample diversity of these examples corroborates the generality of the MAP principle, which becomes even wider as we take into account varying spatial scales and complexity of the agents (e.g.~component, device, machine, production line, whole factory). At the same time, agents can be of several types including mechanical or electronic devices, cells, living beings, factories, programs, etc.

The involved agents can interact regarding the use of the resources in two main ways: independent or dependent. The concepts and methodology developed in the present work focus on the former type of interaction, but some types of dependent operation can also be considered provided they can be associated to the directionality of the resource flow.

\begin{figure}
  \centering
     \includegraphics[width=0.7\textwidth]{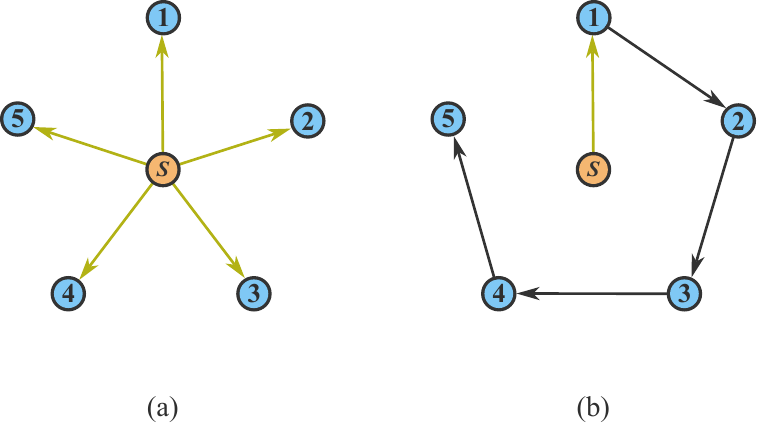}
   \caption{Two main possibilities of supplying resources from a single source into multiple processing agents: (a) parallel; and (b) sequential. The arrows represent the flow of resources (same type or progressively modified) which, typically, is not conserved as a consequence of one or more agents incorporating resources sinking. }\label{fig:ex_net}
\end{figure}

The present work extends and complements a preliminary related work~\cite{da2022dynamics}.
Given that the resources emanate from a single source $S$ (the model also generalizes to a finite number of sources), and that several distinct agents are involved, it becomes necessary to \emph{distribute} the resources among these agents in some specific manner. Figure~\ref{fig:ex_net} illustrates two main possibilities, which are henceforth called \emph{parallel} and \emph{sequential}. 

In the former case, resources are sent directly to each of the involved agent. A completely distinct situation occurs in the latter case, in which the resources are sequentially distributed along the agents. It is also interesting to observe that the directionality of the resources flow between agents, as in Figure~\ref{fig:ex_net}(b), may also indicate \emph{inter-dependence} along the processing, in the sense that the resource to be forwarded by an agent to the next needs to be ready at each discrete time step.  Observe that it is also possible that each of the sequential agents operates independently on the flow of resources, but is nevertheless required to forward a fraction of the received resources to the destination node. At the same time, distribution architectures such as that indicated in Figure~\ref{fig:ex_net}(a) intrinsically assume independence of processing among the agents, as they are not interconnected.

At each discrete time instant $t$, each agent $i$ has a total stored quantity of resources (state) equal to $x_i(t)$, a fraction $s$ of which remains stored into that same agent, a fraction $f$ is forwarded to other agents, and a fraction $e$ is transformed into products. Observe that $s+f+e=1$. Given that each agent may also receive resources, the state of the agent can then be updated as $x_i [t+\Delta t] = s \, x_i[t] + r_i[t]$, where $r_i[t]$ is the total of resources received by agent $i$ from the source and/or other agents. 

The source is henceforth understood to deliver a constant quantity $b$ of resource at each discrete time instant $t=1, 2, \ldots$. For simplicity's sake, we also henceforth understand that the transformed products comprehend both useful and lost components, taking place in respective fractions $u$ and $w$, with $u+w=1$. It is interesting to observe that the fraction $e \, x_i[t]$ can therefore be associated to \emph{resources sinking}.

Interestingly, the eleven considered types of MAP systems can be shown to be characterized by an \emph{steady state}, along which the agents states undergo decreasing asymptotic changes toward the respective equilibrium configuration. The steady state is reached after a respective \emph{transient} state. 

Given a MAP system modeled as described above operating at steady state, several important questions can be addressed, including the following centrally important issues: (i) which resource distribution scheme can lead to maximum production; (ii) how diverse the states of the agents are; and (iii) how long it takes to achieve the steady state.

The present work is aimed at addressing MAP systems whose distribution structure is represented as a network of interconnections between the source and among the agents (e.g.~\cite{albert2002statistical,newman2018networks,costa2007characterization,costa2011analyzing}).
More specifically, eleven types of parallel and sequential MAP architectures have their operation simulated under to parametric configurations corresponding to: (i) keeping most of the resources at each agent; and (ii) forwarding most of the resources to other agents.  Three performance parameters are considered, which quantify the total amount of performed work, the dispersion of state values at the steady state and the transition time from the transient to the steady state regimes. Several interesting results are obtained and discussed.

This work starts by presenting the two basic parallel and sequential MAP architectures, and proceeds by describing the adopted simulation methodology and performance parameters. Then, each of the eleven architectures has its performance estimated and discussed.

\section{Related Works}\label{sec:relatedworks}

As a consequence of the generality of MAP systems, related works have been developed in several areas. This section presents a brief review of some of these works.

Distributed production systems have been typically addressed in Operations Research (e.g.~\cite{koopmans1949optimum, churchman1957introduction, blumenfeld1991synchronizing, hillier2001introduction}), especially in the area of Transportation Theory (e.g.~\cite{jara2007transport, holmberg1999production}), where the distribution from mines to factories is studied with emphasis being placed typically on the transportation cost. Related research has been done also in the area of Supply Chain Management (e.g.~\cite{guinet1991textile, stadtler2014supply, hugos2018essentials, copacino2019supply}), which tends to emphasize the timing, storage, demand, availability, and resilience of materials sources.

The study of MAP systems dynamics is also related to the areas of Dynamical Systems (e.g.~\cite{katok1995introduction}) and Statistical Physics (e.g.~\cite{mandl1991statistical,reichl1999modern,schnoerr2017approximation}). Of particular interest is the concept of \emph{master equation}, which are first order systems of equations modeling probabilistic occupancy of states. The transition probabilities between states can also be related to flow which, in the case of the MAP systems considered in the present work are non-conservative as a consequence of the presence of sources and sinks. Systems involving sources and sinks have received special interest from theoretical ecology and population dynamics (e.g.~\cite{gundersen2001source,eriksson1996regional,crowder2000source,gravel2010source}).

The representation of systems in terms of respective graphs and networks has been typically addressed in the area of Network Science (e.g.~\cite{albert2002statistical,newman2018networks,costa2007characterization,costa2011analyzing}). Particular attention has been placed on the relationship between the topology of interconnections and several types of dynamics in these systems (e.g.~\cite{strogatz2001exploring, boccaletti2006complex, zhang2006complex}).

\section{Basic Distribution Architectures}

\begin{figure}
  \centering
     \includegraphics[width=1\textwidth]{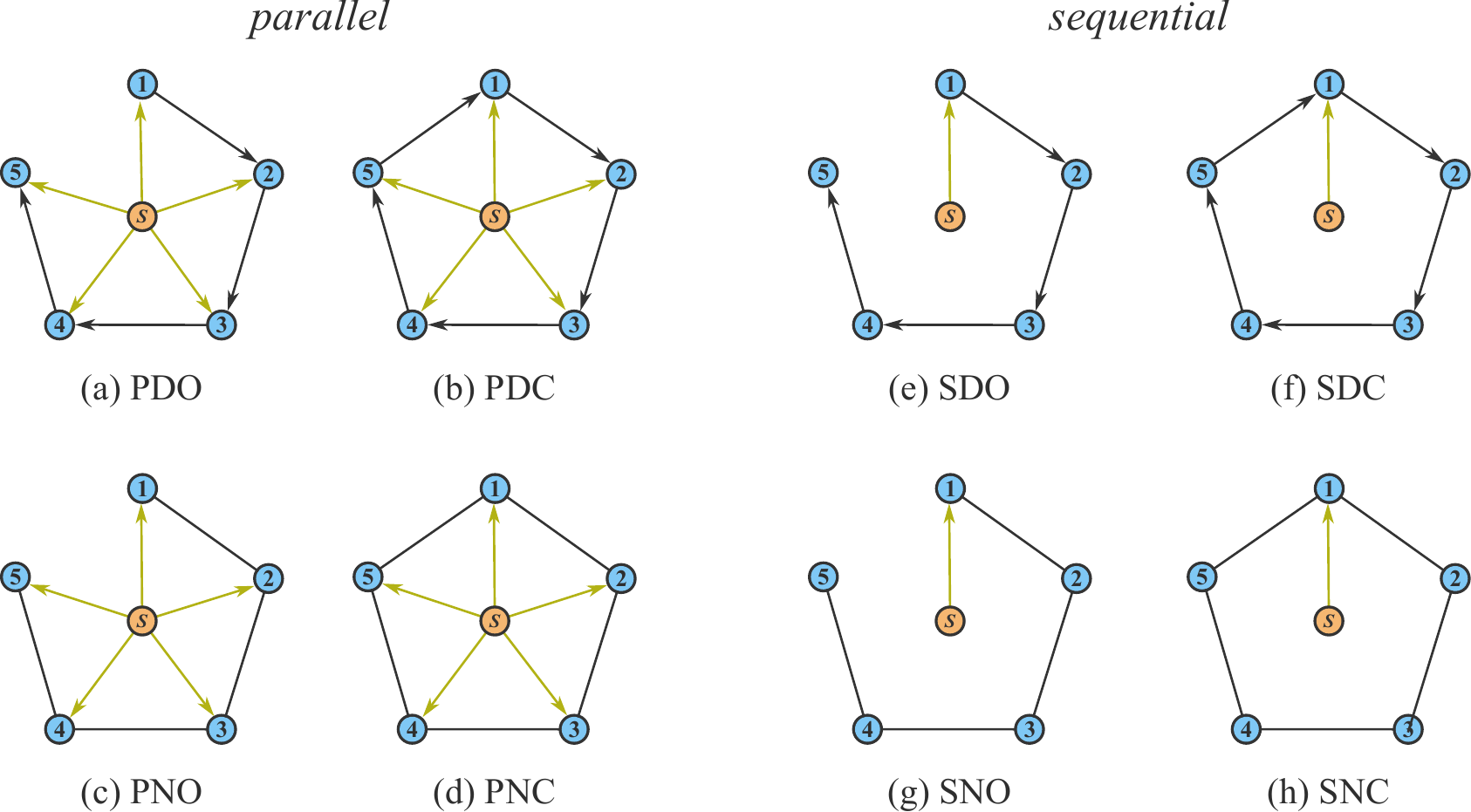}
   \caption{The basic distribution architectures considered in the present work can be subdivided into two main groups: \emph{parallel} and \emph{sequential}. The same type of interconnection between the nodes is adopted respectively to the two groups. Cases (a), (b), (e), and (f) involve \emph{directed} connections between the nodes; being otherwise \emph{undirected} (or bi-directed) in the cases (c), (d), (g) and (h). Though only 5 agents (nodes) are shown in this figure, these eight basic distribution architectures can be readily extended to any number $N$ of agents.}\label{fig:nets}
\end{figure}

In addition to the two reference resource distribution architectures illustrated in Figure~\ref{fig:ex_net}, several other types of architectures are possible. In the present work, we also consider the eight basic architectures cases depicted in Figure~\ref{fig:nets}, which can be organized into two main types of groups: (i) \emph{parallel}, characterized by the source sending resources to all nodes; and (ii) \emph{sequential}, in which the source sends the resource flow only to one of a sequence of nodes. Observe that this classification refers exclusively to the distribution of resources from the source node to agents, and not to the distribution along or among agents. 

The four architectures in each of the two main groups are characterized by the analogous distribution of resources along the agents, with the only difference consisting of the way the resources are provided by the source. The first configuration in each group, namely (a) and (e), is characterized by a direct \emph{open} cycle of resource flow along the agents. The second configurations, namely (b) and (f), involve a closed directed cycle. These two basic configurations taken with non-directed links define the two other configurations in each main group.

We adopt a system with 3 letters to refer to the adopted architectures. The first letter identifies whether the resources are sent to the agents in \emph{parallel} (P) or \emph{sequential} (S) manner. The second letter indicates whether the connections among the agents are \emph{directed} (D) or \emph{non-directed} (N). The last letter states whether the architecture is \emph{open} (O) or \emph{closed} (C).

The choice of the eight adopted distribution architectures has been motivated as a means to study the potential effect of all possible combinations of the three following main important structural aspects: (i) parallel or sequential supply of resources from the source; (ii) directionality of the resource flow along the agents; and (iii) redistribution of resources implemented by closed cycles. More specifically, we will focus on identifying how each of these aspects impacts on the total obtained products and loads on each of the involved agents. Observe that the intrinsic symmetries of configurations (d) and (h) imply respective symmetries in the observed resource dynamics.

\section{Methodology}\label{sec:methodology}

All the considered MAP architectures have their dynamic simulated numerically, involving respective time-discrete master equations. For instance, in the case of the SDO architecture with 5 agents, the following master equation has been employed:
\begin{align}
   \left\{
   \begin{array}{l}
    x_1(t) = s\,x_1(t-1) + b  \\
    x_2(t) = s\,x_2(t-1) + f\,x_1(t-1) \\
    x_3(t) = s\,x_3(t-1) + f\,x_2(t-1) \\
    x_4(t) = s\,x_4(t-1) + f\,x_3(t-1) \\
    x_5(t) = s\,x_5(t-1) + f\,x_4(t-1) \\
   \end{array}
   \right.
\end{align}

As an additional example, the PNC architecture leads to the following master equation:
\begin{align}
   \left\{
   \begin{array}{l}
    x_1(t) = s\,x_1(t-1) + 0.5\,f\,x_2(t) + 0.5\,f\,x_5(t) + 0.2\,b  \\
    x_2(t) = s\,x_2(t-1) + 0.5\,f\,x_1(t) + 0.5\,f\,x_3(t) + 0.2\,b  \\
    x_3(t) = s\,x_3(t-1) + 0.5\,f\,x_2(t) + 0.5\,f\,x_4(t) + 0.2\,b  \\
    x_4(t) = s\,x_4(t-1) + 0.5\,f\,x_3(t) + 0.5\,f\,x_5(t) + 0.2\,b  \\
    x_5(t) = s\,x_5(t-1) + 0.5\,f\,x_1(t) + 0.5\,f\,x_4(t) + 0.2\,b  \\
   \end{array}
   \right.
\end{align}

For reference purposes, Table~\ref{tab:parameters} summarizes the involved dynamic parameters, which are respective to each discrete time step.

\begin{table}[h!]
\centering
\caption{Summary of the parameters controlling the MAP architecture operation.}\label{tab:parameters}
\begin{tabular}{|c||c|}
\hline
\textbf{Parameter} & \textbf{Description}   \\ \hline \hline
$b$       & amount of resources supplied by the source     \\  \hline
$s$       & fraction of resources kept by each agent       \\ \hline
$f$       & fraction of resources forwarded by each agent  \\ \hline
$e$       & \begin{tabular}[c]{@{}c@{}}fraction of resources transformed by each agent \\ into product/work or waste\end{tabular} \\ \hline
\end{tabular}
\end{table}

The equilibrium state of each considered configuration can be verified to be well-defined. It can be readily obtained by solving the respective system of equations imposing $x_i(t+1) = x_i(t)$.

In this work, the analysis of the obtained results will be performed in terms of the three following three performance aspects:

(i) \textit{Total obtained work}: At each time step $t$, for each individual agent $i$, a fraction $e$ of its state $x_i(t)$ is converted into work, or expended as waste. The resulting amount $e \, x_i(t)$ is understood to be subdivided into two respective fractions $w$ and $u$, with $w + u = 1$, so that the amount of obtained work at each time step is $e \, w \, x_i(t)$ and $e \, u \, x_i(t)$. Therefore, along $T$ time steps, the \emph{total amount of work} generated by agent $i$ can be expressed as:
\begin{align}
  W_i = e \, w \, \sum_{k=1}^T x_i(k\, t)
\end{align}

leading to the following total work generated by all the $N$ agents:
\begin{align}
  W_T = \sum_{i=1}^N W(i)
\end{align}

Typically, it is hoped that a given MAP will yield the maximum total work $W_T$ along $T$ time steps. In the present work, we consider $T$ to be large enough so that the MAP system is at the respective steady state. For simplicity's sake and without loss of generality, considering that $w$ is proportional to $e$, we henceforth set $u=0$.

(ii) \textit{The dispersion of the state among the agents}: Along the operation of a MAP system, each agent will possibly have a distinct state $x_i(t)$. The therefore implied load heterogeneity among the agents typically implies the use of distinct agents (differing in the processing capability, and possibly size and cost), or the adoption of identical agents in which case only those operating at the maximum load will be characterized by maximum utilization. These two points indicate that, generally speaking, it is of particular interest to operate with loads as homogeneous as possible. The load dispersion among all the agents in a MAP system is henceforth quantified in terms of the standard deviation of the states $x_i(t)$ after $T$ time steps. 

(iii) \textit{The time it takes to achieve the steady state}: Henceforth, all considered MAP systems initial configuration will be characterized by all agents having null state $x_i(t)$. As the system begins to operate on the received resources, the state at each agent will increase in an asymptotic manner, proceeding through a transient, and then steady state regimes. It can be expected that distinct MAP architectures and configurations will have varying progressions from the former to the latter regimes. In this work, this MAP system property is also considered while taking into account how quickly that system to transition from the transient to the steady states. In this work, we quantify the transition time $\tau$ of a MAP architecture as corresponding to the time it takes from the operation start until the states reach $80\%$ of their equilibrium value. It has been observed that, for the aforementioned MAP architectures, agents sharing a configuration exhibit identical transition times.

\section{Basic Chained MAP Architectures}

\begin{figure}
  \centering
     \includegraphics[width=.9\textwidth]{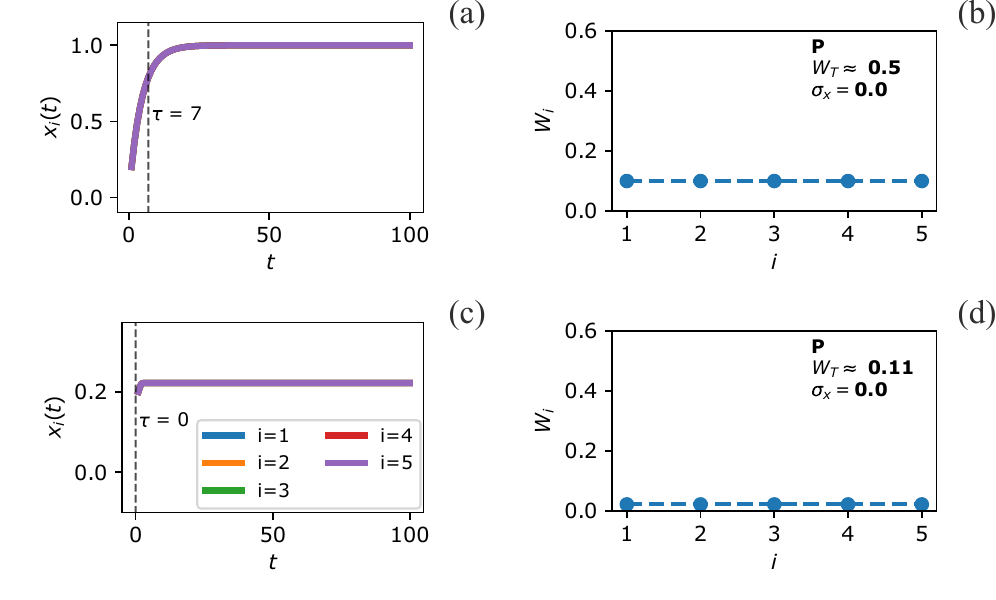}
   \caption{Dynamic characteristics of the basic parallel MAP architecture shown in Fig.~\ref{fig:ex_net}(a) respectively to the two adopted parametric configurations (a) and (b) $s=0.8, f=0.1$, (c) and (d) $s=0.1,f=0.8$. As expected, all agents have identical dynamics. Minimum dispersion $\sigma_x$ is observed in both cases, but this is achieved at the expense of particularly low values of $W_T$. A larger transition time can be observed for the configuration $s=0.8, f=0.1$. }\label{fig:parallel}
\end{figure}

First, we study the basic parallel MAP architecture in Figure~\ref{fig:ex_net}(a), henceforth identified as $P$, respectively to two reference parametric configurations  $s=0.8, f=0.1$ (a) and $s=0.1,f=0.8$ (b). Observe that both these configurations are therefore characterized by $e=0.1$, so that $s+f+e=1$. The obtained results are depicted in Figure~\ref{fig:parallel}.

As could be expected, all five agents, which are all symmetrically identical, undergo identical dynamical characteristics, implying all curves to be superimposed in Figure~\ref{fig:parallel}(a). For the same reason, a null dispersion of resources stored in each agent is also obtained. In addition, because there is no redistribution of resources among the nodes, each node will waste the same fraction $e=f=0.8$ of resources, therefore accounting for the particularly low values of $W_T$ obtained for this architecture respectively to the two considered parametric configurations. The substantially smaller value of $W_T$ observed for the configuration $s=0.1,f=0.8$ is implied by the fact that little amount of resources staying at each node. While near null transition time has been observed for the configuration $s=0.1,f=0.8$, a moderate value $\tau=7$ has been obtained for the other configuration.

The above-discussed results are particularly important in the sense that they indicate that relatively little total work can be obtained by using the basic parallel MAP architecture.  The increase of $W_T$ requires redistribution of resources among the agents, which motivates the other 4 MAP architectures shown in Figure~\ref{fig:nets}. Though other types of interconnectivity between the involved agents are possible, for simplicity's sake we henceforth focus attention only on chained redistribution of resources among agents.

\begin{figure}
  \centering
     \includegraphics[width=.9\textwidth]{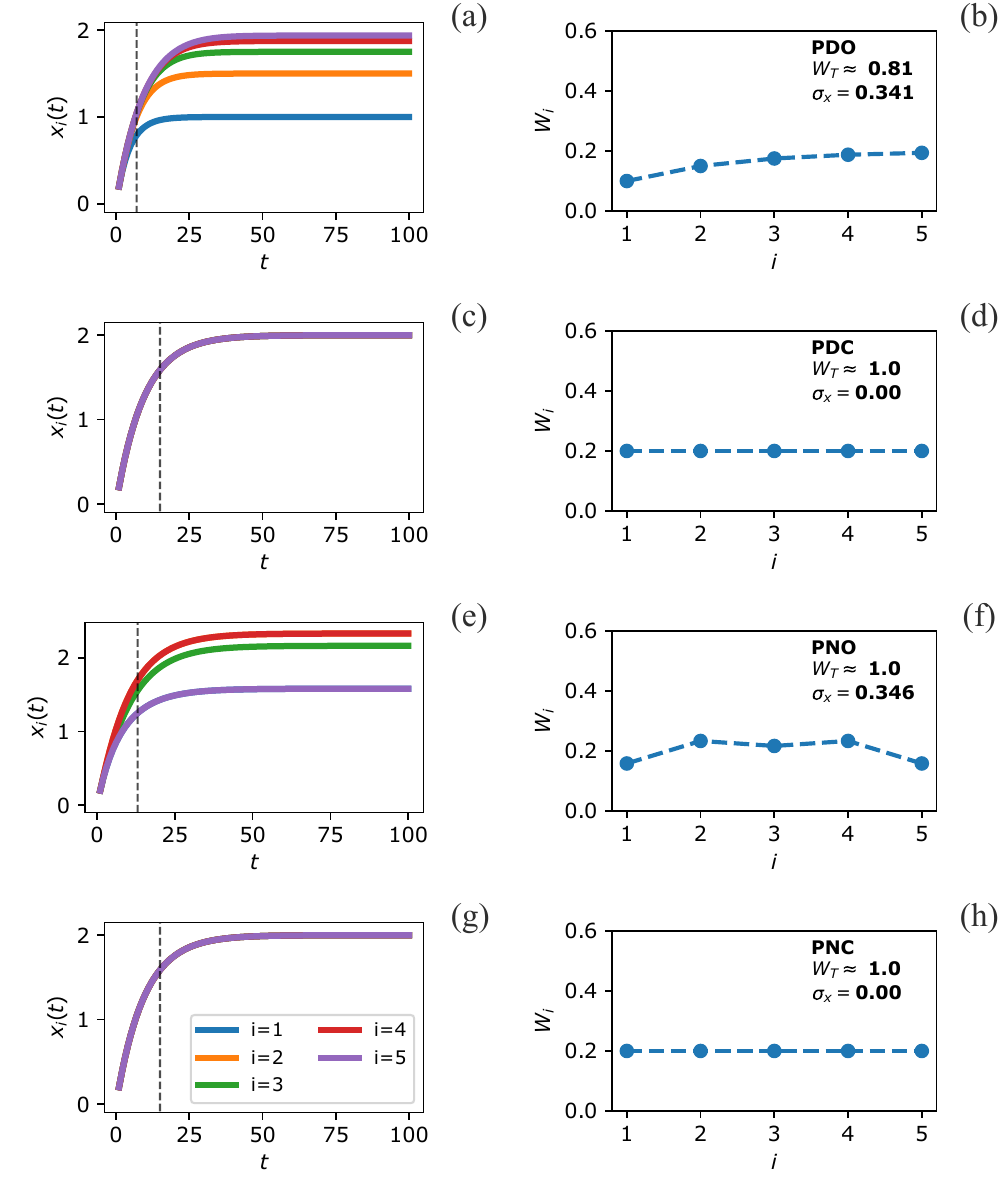}
   \caption{Dynamic properties of the parallel MAP architectures considering the parametric configuration $s=0.8,f=0.1$. The identical interconnectivity between the agents in the PDC and PNC cases is implied in identical dynamics of the agent states, as shown in (c) and (g). These two architectures allowed maximum work $W_T$ and minimum dispersion of state values $\sigma_x$, at the expense of the transition time $\tau$. The dashed lines indicate the respective transition times.}\label{fig:parallel_s}
\end{figure}

\begin{figure}
  \centering
     \includegraphics[width=.9\textwidth]{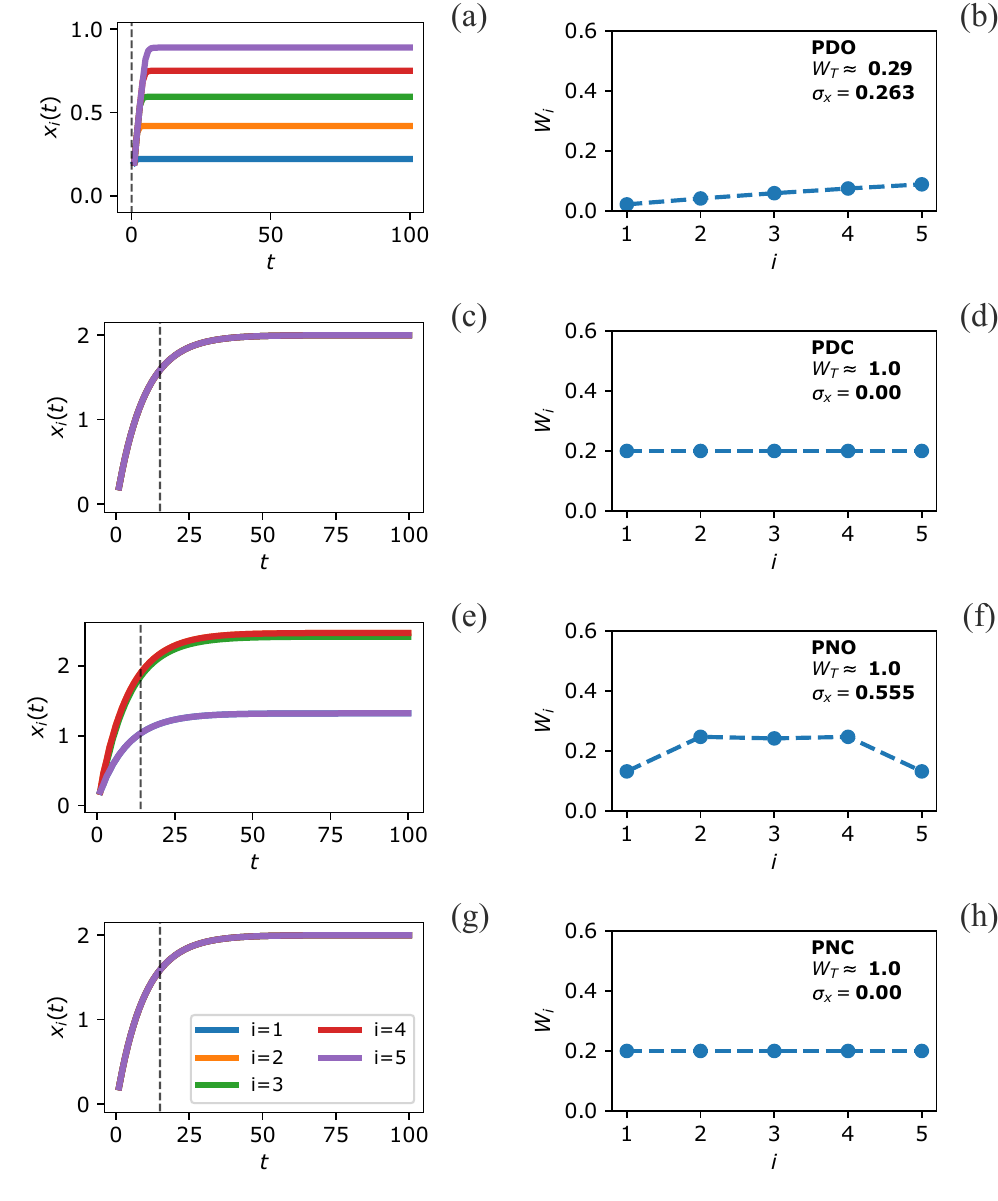}
   \caption{Dynamic properties of the parallel MAP architectures considering the parametric configuration $s=0.1,f=0.8$. As before, the identical interconnectivity between the agents in the PDC and PNC cases is implied in identical dynamics of the agent states, as shown in (c) and (g). These two architectures allowed maximum work $W_T$ and minimum dispersion of state values $\sigma_x$, at the expense of the transition time $\tau$. The main difference between the results obtained by this parametric configuration and those shown in Fig.~\ref{fig:parallel_s} concerns the different state dispersions obtained for the present configuration.}\label{fig:parallel_f}
\end{figure}

\begin{figure}
  \centering
     \includegraphics[width=.9\textwidth]{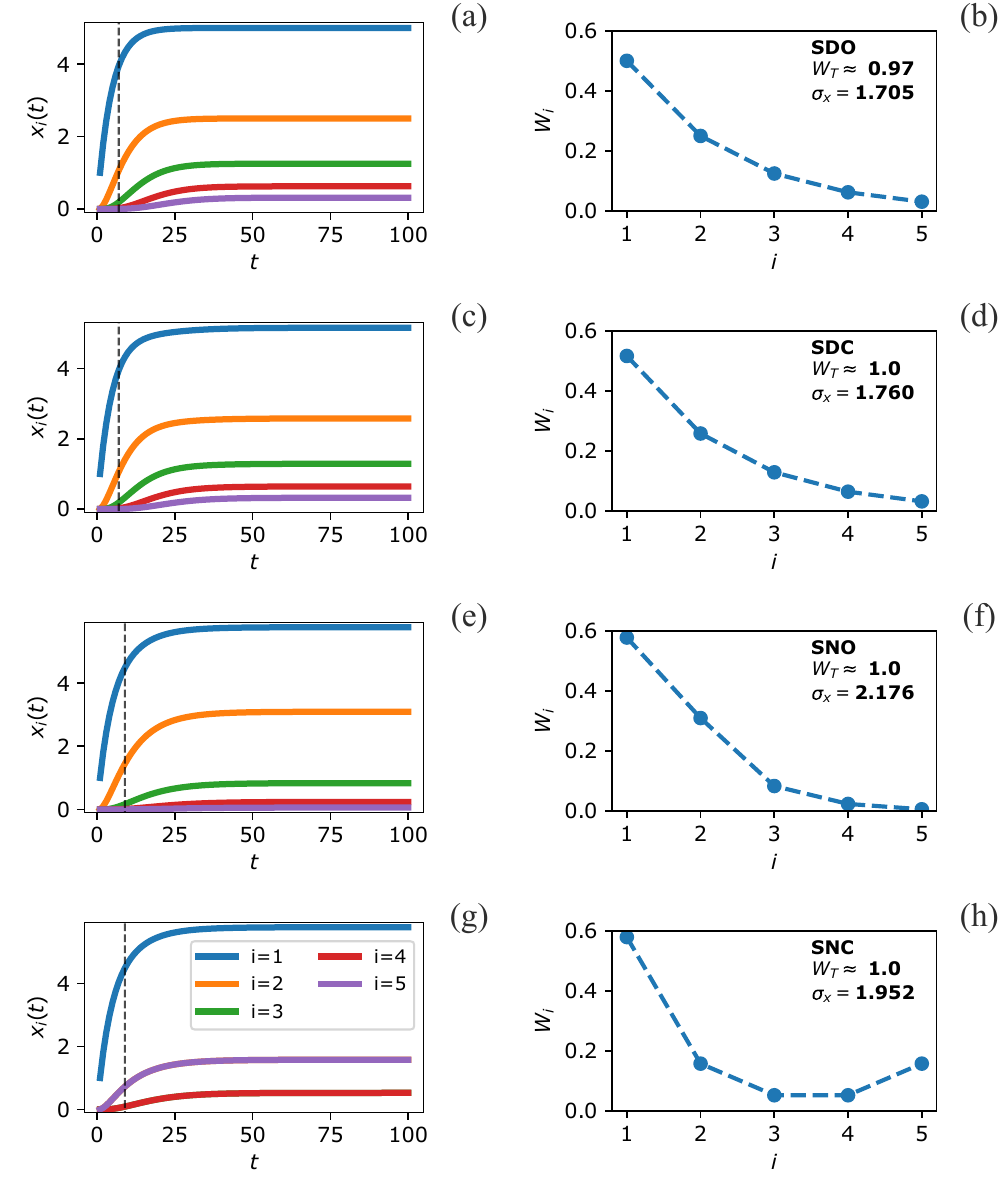}
   \caption{Dynamic properties of the considered sequential MAP architectures respectively to the parametric configuration $s=0.1,f=0.8$. The asymmetry of the agents interconnections implied distinct performances for every considered situation, which is directly reflected in the relatively higher obtained dispersion values $\sigma_x$. Maximum work has been obtained for architectures SDC, SNO, and SNC. The transition times resulted mostly similar.  }\label{fig:sequential_s}
\end{figure}

\begin{figure}
  \centering
     \includegraphics[width=.9\textwidth]{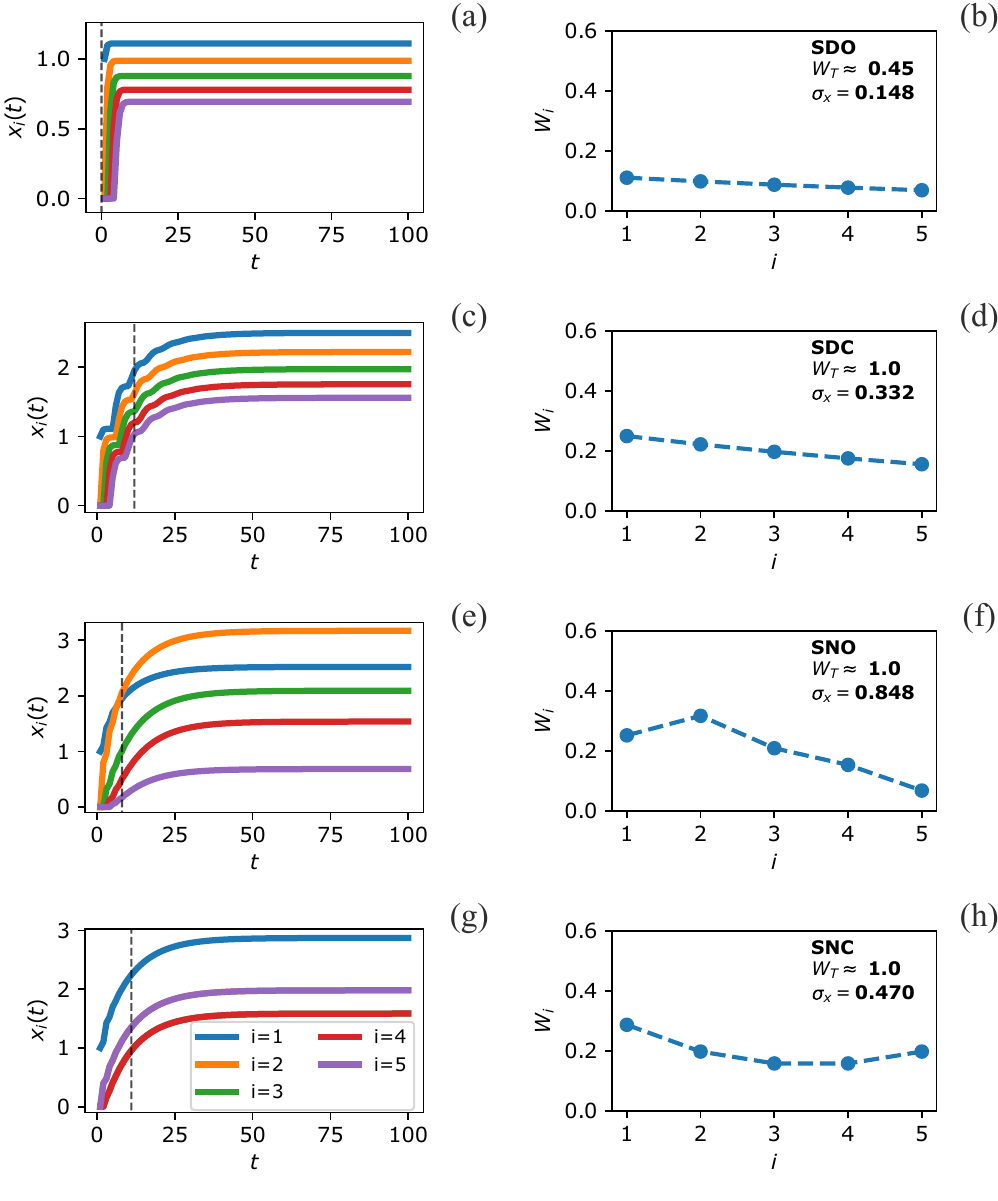}
   \caption{Dynamic properties of the considered sequential MAP architectures respectively to the parametric configuration $s=0.1,f=0.8$. The asymmetry of the agents interconnections implied distinct performances for every considered situation, which is directly reflected in the relatively higher obtained dispersion values $\sigma_x$. Maximum work has been obtained for architectures SDC, SNO, and SNC. Comparatively to the cases shown in Fig.~\ref{fig:sequential_s}, substantially smaller dispersions $\sigma_x$ have been observed for the parametric configuration $s=0.1,f=0.8$.}\label{fig:sequential_f}
\end{figure}

\begin{figure}
  \centering
     \includegraphics[width=.9\textwidth]{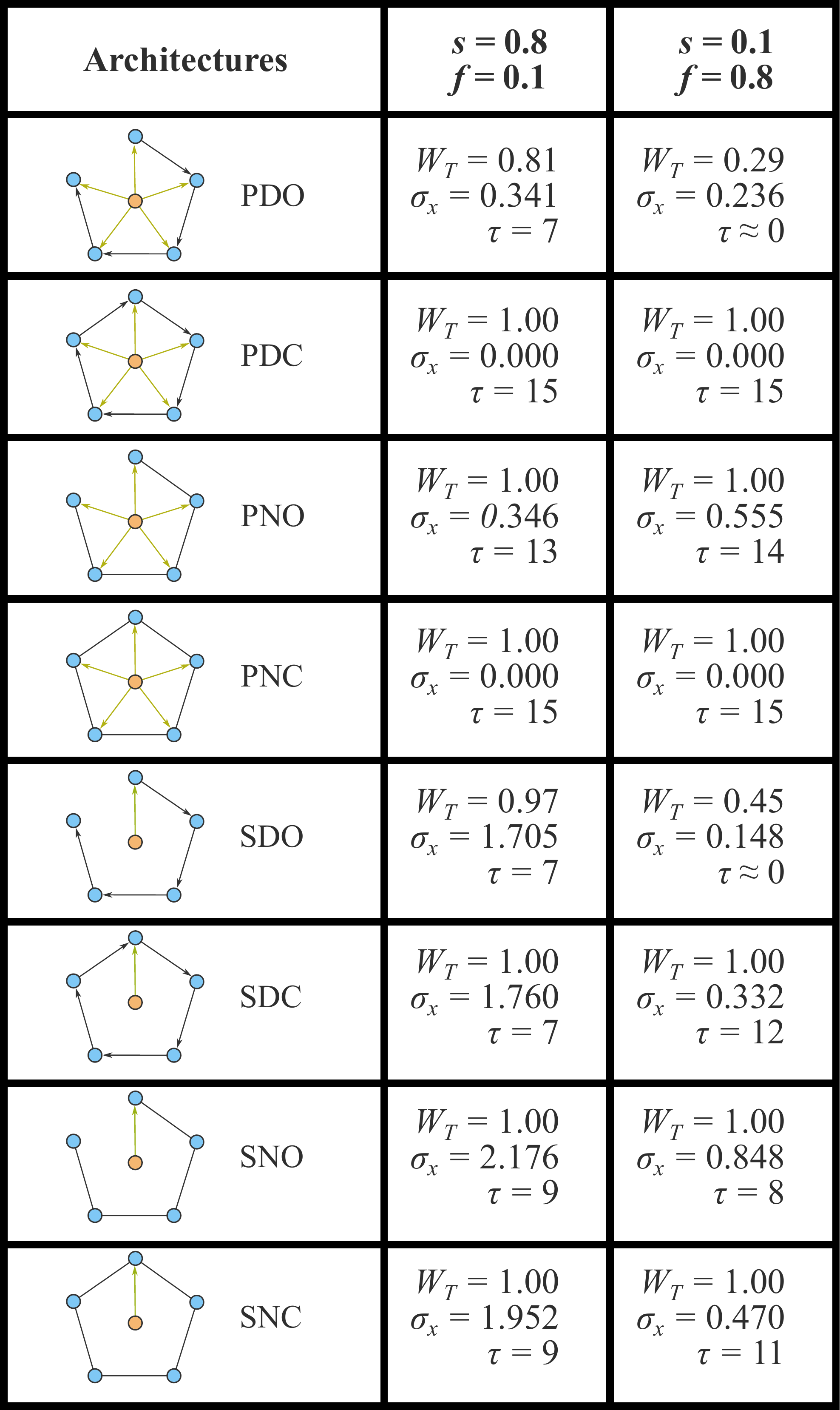}
   \caption{Values of the three performance parameters respectively to each of the considered architectures and the two parametric configurations $s=0.8; f=0.1$ and $s=0.1; f=0.8$. Architectures PDC and PNC allowed the best combination of large total work and small states dispersion for both parametric configurations but at the expense of $\tau$. Null values of $\sigma_x$ have been obtained for PDC and PNC for both configurations. Architectures PDO and SDO allowed near null transition times $\tau$ for the configuration $s=0.1; f=0.8$, but at the expense of $W_T$.}\label{fig:results}
\end{figure}

Figure~\ref{fig:parallel_s} shows the dynamics properties of the considered parallel MAP architectures respectively to the parametric configuration $s=0.8,f=0.1$.  As in Figures~\ref{fig:parallel_f}, \ref{fig:sequential_s}, and \ref{fig:sequential_f} the dynamics of the agents states $x_i(t)$ is shown along the left-hand column, while the total work performed by each agent $W_T$ at the steady state is shown along the right-hand column. Figure~\ref{fig:results} presents a summary of the considered three main performance parameters -- namely $W_T$, $\sigma_x$, and $\tau$.

The unfolding of the state values along time is similar in all cases, involving a transient regime followed by a respective steady state. However, the values of $x_i$ at the steady state, as well as the transition times, resulted specific to each considered architecture. In particular, cases PDC and PNC resulted in identical state values as a consequence of the identical interconnectivity among the agents in those cases. This is not the case for the PDO and PNO architectures, which implied distinct values of $x_i$ at the steady state. The architectures PDC and PNC can be verified to allow maximum work and minimum state values dispersion, but this is achieved at the expense of the highest transition time ($\tau=15$) among the four considered parallel designs, as well as the relatively large number of links required for connecting the source to the agents. The two cases involving open chained interconnection between the agents, namely PDO and PNO, yielded values of $W_T$ that are smaller than the optimal $W_T=1$. That is because the flow of resources is disconnected at the last agent, being therefore wasted.

The features of the dynamics performed by the four basic parallel MAP architectures respectively to the second considered parametric configuration $s=0.1, f=0.8$ is presented in Figure~\ref{fig:parallel_f}.

The obtained results are mostly similar to those respective to the former parametric configuration (Fig.~\ref{fig:parallel_s}), except for the transition times and distinct dispersions observed in the cases PDO and PNO. The PNO case for configuration $s=0.1,f=0.8$ has a greater dispersion in comparison with configuration $s=0.8,f=0.1$ because, for higher values of the forward fraction $f$, the difference between the agents states is intensified (recall that $f=0$ in the architecture P leads to null dispersion). However, the opposite is observed for PNO: the dispersion is reduced with the growth of $f$. In this case, the forward flow of resources is intensified, leading to a larger loss at the last agent, causing the total work ($W_T$), consequently the dispersion ($\sigma_x$), to be reduced.   

In summary, the parallel MAP architectures have been observed to imply a more homogeneous distribution of resources among the agents, to the point that the designs PDC and PNC allow maximum total work and minimum state value dispersions. However, the parallel architectures require a relatively high number of links to be employied while linking the source to the agents.  For these reasons, it becomes interesting to consider also \emph{sequential} MAP architectures, which is done in the following.

Figure~\ref{fig:sequential_s} presents the dynamical performance of the considered sequential MAP architectures respectively to the parametric configuration $s=0.8,f=0.1$.

As could be expected, the sequential (chained) interconnections among the agents implied in completely distinct interconnectivity patterns for each of the agents, with a direct effect in obtaining distinct dynamical performance, including different states $x_i(t)$ being observed along time, as well as relatively higher dispersions $\sigma_x$ resulting for all cases. Similar transition times have been obtained among the sequential designs. Interestingly, maximum total work has been obtained for the SDC, SNO, and SNC MAP architectures. In the case of the SDO design, the waste of resources taking place at the last agent implied $W_T$ to be smaller than 1.

Figure~\ref{fig:sequential_f} depicts the dynamic features of the sequential MAP architectures respectively to the parametric configuration $s=0.1,f=0.8$.

Maximum work resulted for the cases SDC, SNO, and SNC. At the same time, as a consequence of the more intense forward flow of resources, a substantially smaller total amount of work resulted for the architecture SDO. For this same reason, the smallest dispersion ($\sigma_x=0.148$) for the sequential architectures considering $s=0.1,f=0.8$ has been observed also for this architecture. In addition, the SDO architecture also allowed $\tau \approx 0$.

For parallel source distribution, the PDC and PNC provided maximum total work and minimum dispersion at the expense of the transition time respectively to both considered configurations, but at the expense of $\tau$ being the largest among all considered architectures. Among the sequential source distribution architectures in the case, $s=0.8; f=0.1$, the best combination of $W_T$ and $\sigma_x$ has been allowed by the SDO and SDC architectures, with only the latter choice allowing the best-combined performance in the case $s=0.1; f=0.8$. Near null transition times have been observed for the PDO and SDO architectures for the parametric configuration $s=0.1; f=0.8$.

Generally speaking, the MAP architectures which are direct and open necessarily imply a waste of forward resource flow at the last agent, therefore leading to smaller $W_T$. In addition, the dispersion of the state resulted larger for the parametric configuration $s=.1,f=.8$, except for the PNO architecture. A particularly interesting situation has been observed for the SNO architectures respectively to the parametric configuration $s=0.1, f=0.8$, being characterized by the peak of agent work $W_i$ being obtained not for the first agent, as in all other situations, but at the second agent along the resources flow. This is a consequence of the low value $e=0.1$ combined with a particularly intense forward flow $f=0.8$.

\section{All-Connected MAP Architectures}

\begin{figure}
  \centering
     \includegraphics[width=0.7\textwidth]{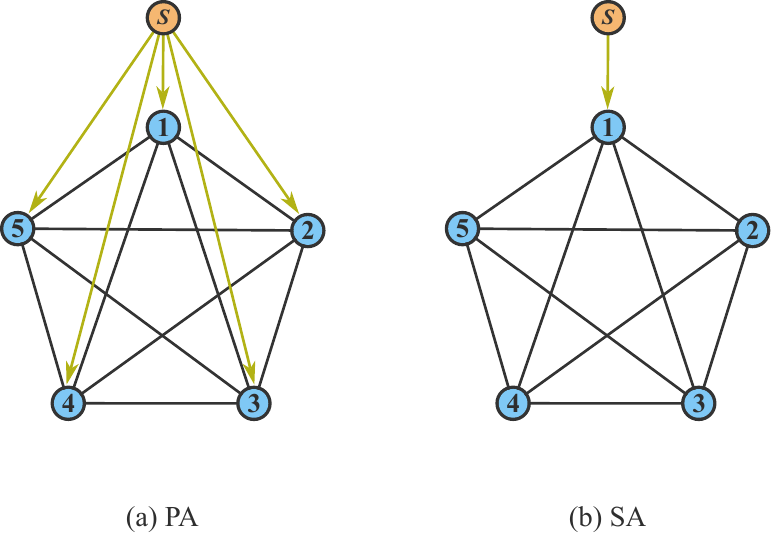}
   \caption{Two MAP architectures PA and SA derived from the basic parallel (P) and sequential cases (SNO), but involving interconnections between all involved agents.}\label{fig:all_connected}
\end{figure}

Considering that all nine MAP architectures analyzed thus far in this study are bound by chained interconnections between agents, it would be worthwhile to investigate further types of interconnections. In the present section, we consider two MAP architectures derived from the basic parallel (P) and sequential (SDO) designs while considering all agents to be interconnected, as illustrated in Figure~\ref{fig:all_connected}. These two architectures are henceforth referred to as PA and SA.

\begin{figure}
  \centering
     \includegraphics[width=.9\textwidth]{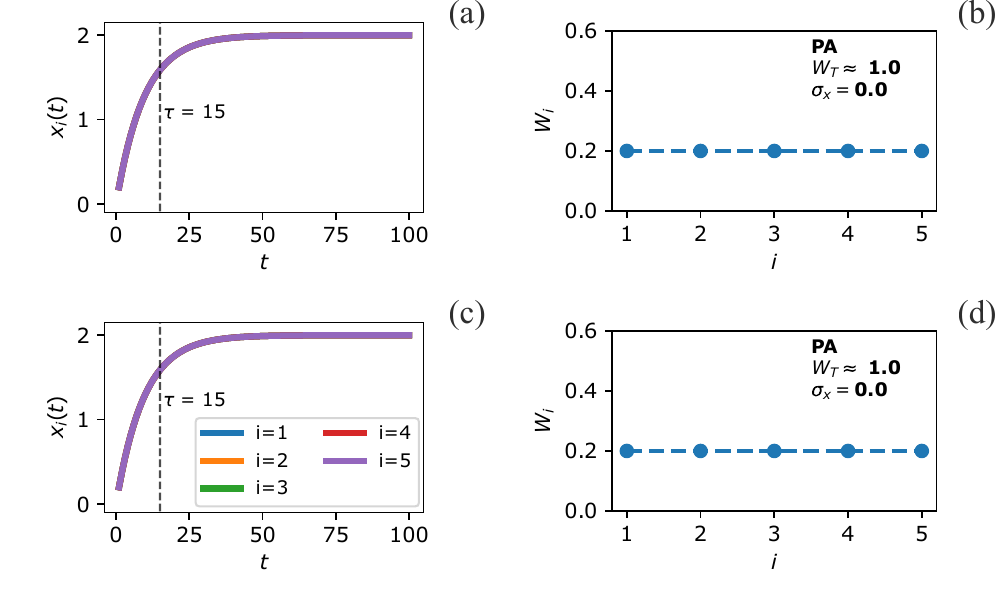}
   \caption{Dynamical properties of the all connected parallel MAP architecture (PA) respectively to two considered parametric configurations $s=0.8; f=0.1$ and $s=0.1; f=0.8$. The identical interconnectivity of each agent leads to identical dynamics. Given that the performance of this architecture is virtually identical to the PDC case shown in Fig.~\ref{fig:parallel_s} and ~\ref{fig:parallel_f}, the additional links between agents required by the PA architecture result can be deemed to be redundant. These architectures are characterized by maximum work and smallest state dispersion, which is achieved at the expense of the transition time $\tau$.}\label{fig:PA}
\end{figure}

The architectures PA and SA are special in the sense of providing the maximum interconnectivity between the involved agents, and therefore maximum redistribution of resources received from the source. Figure~\ref{fig:PA} illustrates the dynamical performance of the PA architecture respectively to the parametric configurations $s=0.8, f=0.1$ and $s=0.1, f=0.8$.

As a consequence of the identical connections of each agent, all state values $x_i$ resulted in identical along time, while identical works $W_i$ were performed by each agent. Therefore, this architecture allows maximum total work $W_T=1.0$ and minimal dispersion $\sigma_x=0$, which is achieved at the expense of a substantially large $\tau=15$. The MAP architectures PDC and PNC achieved equivalent performance with significantly fewer interconnections, making them more efficient. In summary, we have that the inclusion of a considerable number of links between the agents in the PA design did not contribute to improving any of the considered performance parameters.
 
\begin{figure}
  \centering
     \includegraphics[width=.9\textwidth]{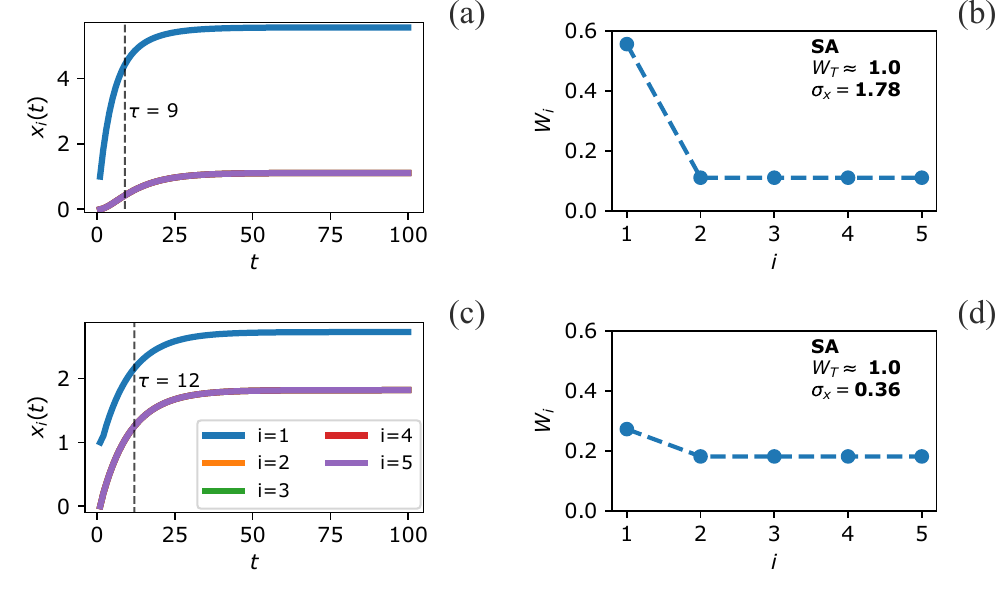}
   \caption{Dynamical properties of the all connected sequential architecture (SA) respectively to two considered parametric configurations $s=0.8; f=0.1$ and $s=0.1; f=0.8$. The identical interconnectivity of agents 2,3,4, and 5 implied in a respective group of state values $X_i(t)$ (shown in violet), while the agent receiving resources directly from the source defined an individual curve (shown in blue). The dispersions of states $\sigma_x$ can be verified to be smaller than those allowed by the architectures SDC and SNC.}\label{fig:SA}
\end{figure}

The dynamical performance of the SA architecture is presented in Figure~\ref{fig:SA} respectively to the parametric configurations $s=0.8, f=0.1$ and $s=0.1, f=0.8$.

Unlike the parallel group of MAP architectures observed earlier, the all interconnected sequential design led to better dispersions of $\sigma_x$ when compared to the previous SDC and SNC architectures. Actually, only the node that directly received resources from the source had a distinct state value unfolding $x_1(t)$ and performed work $W_1$. Therefore, among the sequential designs considered, the SA architecture achieved the best combination of total work and state value dispersion.

\begin{figure}
  \centering
     \includegraphics[width=.7\textwidth]{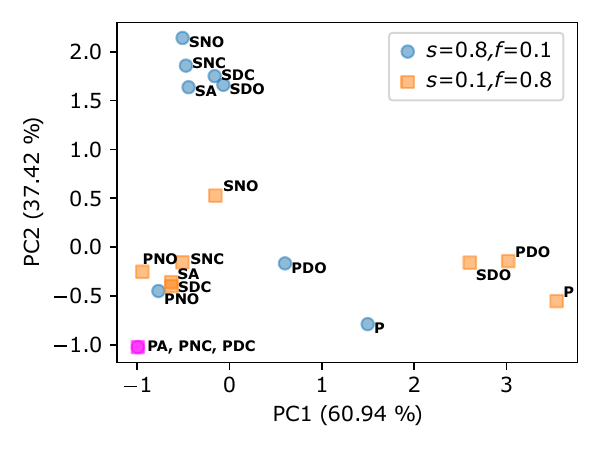}
   \caption{Principal component analysis of the eleven MAP architectures considered in this work respectively to both parametric configurations $s=0.8,f=0.1$ (circle, blue) and $s=0.1,f=0.8$ (square, orange), considering the three performance parameters as data. Of particular interest is the concentration of sequential designs at the upper left-hand corner of the plot. Observe also the superimposition onto a single point (in magenta) of the PA, PNC, and PDC architectures for both parametric configurations. The variance explanation accounted for by each axis is respectively indicated.}\label{fig:PCA}
\end{figure}

Figure~\ref{fig:PCA} presents a Principal Component Analysis (PCA, e.g.~\cite{jolliffe2011principal, gewers2018principal}), considering the three performance indices ($W_T$, $\sigma_x$, and $\tau$), respective to the 11 MAPs considered in this study, including the two parameter configurations ($s=0.8,f=0.1$ and $s=0.1,f=0.8$).

First, we have that the two first principal components accounted for 98.36 \% of the total data variance, therefore indicating a good data representation by the obtained projection. Interestingly, the sequential architectures for the parametric configuration $s=0.8,f=0.1$ resulted in a relatively compact group at the upper left-hand corner of the plot. This indicates that these cases have markedly similar performance despite the fact of having distinct interconnections between the respective agents. Identical performances have been observed for the PA, PNC, and PDC respectively to both considered parametric configurations.

\section{Concluding Remarks}

The MAP paradigm can be used to analyze, model, and optimize a wide range of real-world and abstract situations involving the transformation, by multiple agents, of a given resource into respective products. In the present work, we described a systematic quantitative analysis of the performance of eleven basic MAP architectures, which have been subdivided into two main groups according to the manner in which the resources are supplied from a single source to multiple agents. The several considered designs in each of these two main groups differ from one another regarding the type of interconnection between the involved agents.

The dynamics of the considered MAP architectures involved, at each subsequent discrete time step, each agent keeping a fraction $s$ of the resources, forwarding a fraction $f$, and expending a fraction $e = 1 - s-f$ to produce work (transform the resource into some product). The dynamic of each MAP configuration starts with every agent having a null amount of resources and proceeds along a transient regime, followed by a respective steady state regime defined by the respective equilibrium state.
The performance of each MAP architecture has been quantified in terms of the total quantity of produced work $W_T$, the dispersion of the agents state values $\sigma_x$, as well as the transient time $\tau$.
Though the reported study involves 5 agents, similar results could be expected for systems involving larger numbers of agents. Several interesting results have been obtained, as summarized in the following respectively to the three considered performance parameters.

The total performed work $W_T$ obtained from a given MAP architecture has been found to be mostly related to the amount of resources that are wasted at the last agent along the flow of resources. Thus, except for the open designs PDO and SDO, which are involve resources loss, all other considered architectures allowed maximum total work $W_T=1.0$ for both considered parametric configurations $s=0.8, f=0.1$ and $s=0.1, f=0.8$. Between the two architectures with suboptimal total work, both resulted in less effective performed work respectively to the parametric configuration $s=0.1, f=0.8$, which involves a more intense forwarded resource flow.

Minimum dispersion of state values among the agents ($\sigma_x=0$) has been observed for both parametric configurations respectively to the P, PDC, PNC, and PA architectures, all of which belong to the considered parallel group of MAP architectures. That is a direct consequence of the identical pattern of interconnection between all the involved agents. 

Among the sequential architectures, the smallest value of $\sigma_x=0.148$ was obtained for the SDO with the configuration $s=0.1, f=0.8$. This result is particularly surprising, especially when compared to the SA architecture because it involves the smallest number of interconnections among all considered MAP architectures. The SDC architecture also allowed the smallest transition time. However, these two interesting performance parameters have been obtained at the expense of the total performed work, which was only $W_T=0.45$. The MAP architecture with the second smallest value of $\sigma_x=0.332$ was allowed by the SDC design with the parameter configuration $s=0.1, f=0.8$.  This same case was also characterized by maximum performed work $W_t=1.0$ and moderate transition time $\tau=12$. As such, the SDC architecture operating with $s=0.1, f=0.8$ represents an interesting overall alternative respectively to a combination of the three considered performance parameters. Another interesting result concerns the fact that the architecture SA with $s=0.1, f=0.8$, allowed only the third smallest dispersion despite its substantial number of interconnections between agents.

Though we have focused on a single transient state unfolding from a null agent state, the respective transition times can still provide insights also regarding the MAP architectures in alternative situations involving variable resource flow emanating from the source. More specifically, an architecture with large value of $\tau$ is likely to adapt slowly to flow variations, while a design with a small $\tau$ will tend to adapt more quickly.

Another particularly interesting result discussed in the present work concerns the fact that the consideration of the parallel design involving all interconnected agents (PA) did not contribute, in general, to improving the considered performance parameters.

All in all, neither of the eleven considered MAP architectures allowed optimal performance regarding all the three parameters $W_T$, $\sigma_x$, and $\tau$, with improvement in one or two of these parameters being typically achieved at the expense of the remainder performance parameters. For instance, the PDO architecture with parametric configuration $s=0.1, f=0.8$ allows small $\tau$ and $\sigma_x$ at the expense of $W_T$. The PDC design for this same parametric configuration was observed to yield high $W_T$ and small $\sigma$, but at the expense of implying the largest $\tau$. In the case of the parametric configuration $s=0.8, f=0.1$, the SDC architecture yielded maximum $W_T$ at the expense of $\sigma_x$.

The generality of the concepts, methods, and results described and discussed in the present work paves the way to a number of interesting further research. This includes the consideration of other MAP architectures involving networks of interconnections between the respective agents, as well as the consideration of other parametric configurations. It would also be of particular interest to study real-world systems from the perspective of the approaches described here.

\section*{Acknowledgments}
Alexandre Benatti thanks MCTI PPI-SOFTEX (TIC 13 DOU 01245.010222/2022-44).
Luciano da F. Costa thanks CNPq (grant no.~307085/2018-0) and FAPESP (grants 15/22308-2 and 2022/15304-4).

\bibliography{ref}
\bibliographystyle{unsrt}

\end{document}